\documentstyle[preprint,aps]{revtex}

\title{Conservation laws preserving algorithms for
       spin dynamics simulations}

\author{I. P. Omelyan,$^1$ I. M. Mryglod,$^{1,2}$ and R. Folk$^2$}

\address{$^1$Institute for Condensed Matter Physics,
         1 Svientsitsky Str., UA-79011 Lviv, Ukraine}
\address{$^2$Institute for Theoretical Physics, University of Linz,
         A-4040 Linz, Austria}
\date{\today}

\begin{document}

\maketitle
\begin{abstract}

We propose new algorithms for numerical integration of the equations
of motion for classical spin systems with fixed spatial site positions.
The algorithms are derived on the basis of a mid-point scheme in
conjunction with the multiple time staging propagation. Contrary
to existing predictor-corrector and decomposition approaches, the
algorithms introduced preserve all the integrals of motion inherent
in the basic equations. As is demonstrated for a lattice ferromagnet
model, the present approach appears to be more efficient even over
the recently developed decomposition method.

\end{abstract}


The prediction of collective phenomena in magnetic materials was
the subject of many investigations in theory and computer experiment
\cite{Watson,Amit,Parisi,Kob,Tavar,Nijmei}. These investigations dealt
mainly with static properties, in particular, with phase transitions,
critical behavior and scaling. The theoretical description of dynamic
properties presents a much more difficult problem which, at the present
time, cannot be solved quantitatively even for the simplest lattice
systems such as the Ising, XY, and Heisenberg models. Until now, the
method of molecular dynamics (MD) can be considered as the main tool
for quantitative studies of dynamic critical behavior, dynamic scaling,
and processes of spin relaxation.

The construction of stable and efficient MD algorithms for systems with
spin (orientational) degrees of freedom remains a current problem. The
traditional numerical methods \cite{Burden} on integrating differential
equations are unsuitable because of their high instability on MD scales
of time. For this reason, standard predictor-corrector schemes were
utilized in recent years for MD simulations of classical XY and Heisenberg
models in $d = 2$ and of Heisenberg ferro- and antiferromagnets in $d = 3$
dimensions \cite{ChenLan,EvLan,BunChen,CosLan,LanKre}. However, the
requirement on total energy conservation restricts these schemes to
be used with very little step sizes only.

Quite recently, new spin dynamics algorithms have been devised
\cite{Krech,Landau,Tsai}. They are based on the Suzuki-Trotter (ST)
decompositions of exponential operators and unlike usual methods preserve
the total energy to within machine accuracy. It was shown that these
algorithms allow much larger time steps than the predictor-corrector
schemes and thus may lead to a substantial speedup of MD simulations.
However, the decomposition integrators destroy the conservation of
magnetization vector. Moreover, they are applicable, in fact, only to
spin systems when the decomposition into noninteracting sublattices is
possible and cannot be used for models with arbitrary lattice structures.

In this report we solve the problems just mentioned and derive algorithms
which preserve all the conservation laws imposed by the equations of
motion. The algorithms are tested and compared with previous approaches.

Let us consider a collection of $N$ spins represented by continuous
three-component vectors ${\bf s}_i=(s_i^x, s_i^y, s_i^z)$ with the fixed
length $|{\bf s}_i| = 1$ for each site $i$. A typical model Hamiltonian
for such a system can be cast in the form
\begin{equation}
H = - \sum_{i<j}^N J_{ij} \big( s_i^x s_j^x + s_i^y s_j^y + \lambda
s_i^z s_j^z \big) - C \sum_{i=1}^N (s_i^z)^2
\end{equation}
where $J_{ij}$ is the exchange integral for a pair $(i,j)$ of spins,
$\lambda$ is the exchange anisotropy parameter, and $C$ denotes the
strength of single-site field anisotropy. At $C=0$, Eq. (1) represents
the isotropic ($\lambda=1$) or anisotropic ($\lambda \ne 1$) Heisenberg
ferro- or the corresponding antiferro-magnet for $J > 0$ and $J < 0$,
respectively. For $\lambda = C = 0$, Eq. (1) reduces to the XY model.
We do not restrict ourselves to lattice systems with the nearest-neighbor
interaction, and the results presented below can be used for continuum
models with arbitrary spatial spin distributions as well. Therefore, we
indicate explicitly by the subscripts $i,j$ that the exchange integral
$J_{ij}$ depends on spatial positions (which are fixed but not
necessarily periodic) of spin sites.

The dynamic properties of the system can be obtained from MD
simulations by numerical integrating the following equations of
motion \cite{ChenLan,EvLan,BunChen,CosLan,LanKre,Krech}:
\begin{equation}
{\bf \dot s}_i \equiv \frac{\rm d}{{\rm d} t} {\bf s}_i = \frac{1}{\hbar}
\frac{\partial H}{\partial {\bf s}_i} {\mbox{\boldmath $\times$}}
{\bf s}_i(t) \equiv {\bf \Omega}_i(t) {\mbox{\boldmath $\times$}}
{\bf s}_i(t) ,
\end{equation}
where ${\bf \Omega}_i=-\frac{1}{\hbar} \big( \sum_{j(j \ne i)} J_{ij}
(s_j^x, s_j^y, \lambda s_j^z) + 2 C (0,0,s_i^z) \big)$ denotes the local
Larmor frequency. Since the effect of collective thermal excitations
(e.g., phonons) is not described by the Hamiltonian (1), Monte-Carlo (MC)
simulations must be employed additionally \cite{Krech} to generate
equilibrium configurations as initial conditions to Eq. (2). This
procedure is justified by the fact that in magnetic systems the
characteristic time intervals corresponding to varying spin variables
are much shorter than typical time scales of the thermal excitations.

In view of the symmetry $J_{ij}=J_{ji}$, it follows from Eqs. (1) and (2)
that the total energy $E \equiv H$ is an integral of motion, i.e. ${\rm d}
E / {\rm d} t = 0$. The magnetization ${\bf M} = \sum_i {\bf s}_i$ is also
conserved during the spin evolution of the isotropic Heisenberg model.
For the anisotropic case ($\lambda \neq 1$ and/or $C \neq 0$) only the
component $M_z$ of ${\bf M}$ will unchange in time. In addition, the
structure of Eq. (2) imposes also the conservation of individual spin
lengths. Existing MD algorithms do not fulfill these conservation laws
simultaneously. Thus, in order to reproduce the dynamical behavior
properly it is required that the deviations of conservative quantities
from their exact values to remain within an acceptable level of precision.
This leads to obvious limitations on the size of time steps which can be
used in MD simulations. It would, therefore, be very desirable to derive
algorithms which conserve all the integrals of motion exactly or, at
least, within machine accuracy.

The basic idea of our approach consists in the following. Suppose that an
initial spin configuration $\{ {\bf s}_i(t) \}$ has been specified and we
would like to obtain values of ${\bf s}_i$ at time $t+\tau$ within ${\cal
O}(\tau^3)$ truncation terms, where $\tau$ denotes the step size. This can
be realized using a mid-point scheme, ${\bf s}_i(t+\tau)={\bf s}_i(t)+{\bf
\dot s}_i(t+\tau/2) \tau + {\cal O}(\tau^3)$. The time derivative can be
determined applying the usual interpolation formula ${\bf \dot s}_i(t+
\tau/2)=\frac12 [{\bf \dot s}_i(t)+{\bf \dot s}_i(t+\tau)] + {\cal O}
(\tau^2)$. Such a formula, however, does not maintain the unit norm of
${\bf s}_i(t+\tau)$ and thus needs in modifications. Since ${\bf \dot s}_i$
depends on both the local frequency ${\bf \Omega}_i$ and spin orientation
${\bf s}_i$, it is more natural to apply the interpolation with respect to
these two dynamical variables separately rather than to the function ${\bf
\dot s}_i$ as a whole. In doing so we obtain ${\bf \dot s}_i(t+\tau/2)=
\frac14 [({\bf \Omega}_i(t)+{\bf \Omega}_i(t+\tau)) {\mbox{\boldmath
$\times$}} ({\bf s}_i(t)+{\bf s}_i(t+\tau))]+{\cal O}(\tau^2)$, resulting
in the implicit spin propagation
\begin{eqnarray}
{\bf s}_i^{(n+1)}(t+\tau)&=&{\bf s}_i(t) + \frac{\tau}{2}
\big[\, \tilde{\bf \Omega}_i^{(n)} {\mbox{\boldmath $\times$}}
\big( {\bf s}_i(t) + {\bf s}_i^{(n)}(t+\tau) \big) \, \big]
\nonumber \\ &+& {\cal O}(\tau^3) ,
\end{eqnarray}
where $\tilde{\bf \Omega}_i^{(n)}=\frac12 [{\bf \Omega}_i(t) + {\bf
\Omega}_i^{(n)}(t+\tau)]$. As far as the mid-step frequency $\tilde{\bf
\Omega}_i^{(n)}$ depends itself on (generally speaking) all spin
orientations at time $t+\tau$, Eq. (3) constitutes a set of $N$ quadratic
equations for $\{ {\bf s}_i(t+\tau) \}$ which can be solved iteratively
($n=0,1,2,\ldots$) putting ${\bf s}_i^{(0)}(t+\tau)={\bf s}_i(t)+{\bf
\Omega}_i(t) {\mbox{\boldmath $\times$}} {\bf s}_i(t) \tau$ as the initial
guess (note that iterative solutions are inherent in predictor-corrector
and decomposition schemes too). Already one iteration is enough to reach
the required ${\cal O}(\tau^3)$ accuracy. The necessity of performing
further several updates of Eq. (3) will be motivated latter.

It can be shown readily that for the isotropic model the magnetization is
conserved exactly during the spin dynamics propagation given by Eq. (3).
Indeed, summation of Eq. (3) over the spin numbers and taking into account
the explicit expression for $\tilde{\bf \Omega}_i^{(n)}$ yields ${\bf
M}^{(n+1)}(t+\tau)={\bf M}(t)+\Delta{\bf M}^{(n)}$, where $\Delta{\bf
M}^{(n)} = \frac{\tau}{4 \hbar} \sum_{i \ne j} J_{ij} [{\bf s}_i^{(n)}(t)
+ {\bf s}_i^{(n)}(t+\tau)] {\mbox{\boldmath $\times$}} [{\bf s}_j^{(n)}(t)
+ {\bf s}_j^{(n)}(t+\tau)]$. The term $\Delta{\bf M}^{(n)}$ is canceled
because of the invariance of the double sum with respect to the
transformation $i \leftrightarrow j$, and of the obvious equality ${\bf a}
{\mbox{\boldmath $\times$}} {\bf b} + {\bf b} {\mbox{\boldmath $\times$}}
{\bf a}=0$ which is valid for arbitrary vectors ${\bf a}$ and ${\bf b}$.
Thus, ${\bf M}(t+\tau)={\bf M}(t)$ within each iteration. The proof of the
conservation $M_z(t+\tau)=M_z(t)$ for the anisotropic case is similar.

Another important feature of the mid-point propagation is that it conserves
the total energy within machine accuracy ${\cal O}(\varepsilon)$, where
$\varepsilon$ denotes the iterative precision, i.e., $|{\bf s}_i^{(n+1)}
(t+\tau)-{\bf s}_i^{(n)}(t+\tau)| < \varepsilon$. To show this, let us
perform a scalar multiplication of Eq. (3) with the vector $\tilde{\bf
\Omega}_i=\frac12 [{\bf \Omega}_i(t) + {\bf \Omega}_i(t+\tau)]$. Then
using the equality $[{\bf a} {\mbox{\boldmath $\times$}} {\bf b}] {\mbox
{\boldmath $\cdot$}} {\bf a} = 0$ one obtains
$$
{\bf s}_i(t+\tau) {\mbox{\boldmath $\cdot$}}
[{\bf \Omega}_i(t) + {\bf \Omega}_i (t+\tau)] =
{\bf s}_i(t) {\mbox{\boldmath $\cdot$}}
[{\bf \Omega}_i(t) + {\bf \Omega}_i (t+\tau)] ,
$$
where ${\cal O}(\varepsilon)$ terms have been neglected. Summing up the
last relation leads to $E(t+\tau) \equiv \frac{\hbar}{2} \sum_i {\bf
s}_i(t+\tau) {\mbox{\boldmath $\cdot$}} {\bf \Omega}_i(t+\tau) = \frac
{\hbar}{2} \sum_i {\bf s}_i(t) {\mbox{\boldmath $\cdot$}} {\bf \Omega}_i
(t)+\Delta E$, where $\Delta E = \frac{\hbar}{2} \sum_i [{\bf s}_i(t)
{\mbox{\boldmath $\cdot$}} {\bf \Omega}_i (t+\tau) - {\bf s}_i(t+\tau)
{\mbox{\boldmath $\cdot$}} {\bf \Omega}_i (t)]$. The term $\Delta E$ is
canceled again because of the linear dependency of ${\bf \Omega}_i$ on
spin components, and of the symmetry $J_{ij}=J_{ji}$, so that $E(t+\tau)=
E(t)+{\cal O}(\varepsilon)$. The uncertainty $\varepsilon$ can be reduced
to a negligibly small value at a given $\tau$ by adjusting the number
$l>1$ of iterations for Eq. (3). The rapid convergence $\varepsilon \to
+0$ is guaranteed by the power dependence $\varepsilon \sim {\cal O}
(\tau^{l+2})$ and by the smallness of $\tau$. Of course, the iterative
solutions require additional computational efforts, but they are
compensated completely by using larger time steps. For instance,
spending the same amount of computer time, we could try to reduce the
energy deviations within only one iteration by decreasing the time step
to $\tau/l$. This way, however, is very inefficient because then the
deviations will behave like ${\cal O}((\tau/l)^3)$ and, thus, decrease
with increasing $l$ much more slower than the power dependence ${\cal
O}(\tau^{l+2})$, in other words $\tau^{l+2} \ll (\tau/l)^3$.

An additional surprising property of the mid-point integration is the
conservation of spin lengths. Implicit evaluations given by Eq. (3) achieve
this conservation in iterative sense, i.e., $|{\bf S}_i(t+\tau)|=1+{\cal
O}(\varepsilon)$. In order to maintain spin lengths exactly, the iteration
process should be reconstructed. Considering the quantity $\tilde{\bf
\Omega}_i^{(n)}$ as a parameter, Eq. (3) can be solved analytically,
\begin{eqnarray}
{\bf s}_i^{(n+1)}&&(t+\tau) = \frac{1}{1+\frac{\tau^2}{4}
\big(\tilde{\Omega}_i^{(n)}\big)^2} \Big[ {\bf s}_i(t) +
\tilde{\bf \Omega}_i^{(n)} {\mbox{\boldmath $\times$}}
{\bf s}_i(t) \tau \nonumber  \\
&& + \frac{\tau^2}{4} \Big( 2 \, \tilde{\bf \Omega}_i^{(n)}
\big( \tilde{\bf \Omega}_i^{(n)} {\mbox{\boldmath $\cdot$}} \,
{\bf s}_i(t)) - \big(\tilde{\Omega}_i^{(n)}\big)^2
{\bf s}_i(t) \big) \Big) \Big] ,
\end{eqnarray}
and further iterated because of the explicit dependence of $\tilde{\bf
\Omega}_i$ on spin orientations $\{ {\bf s}_i(t+\tau) \}$. Obviously,
such a modified iterative scheme will conserve the magnetization and
total energy (to within machine accuracy) like the usual spin evaluation
(3) (since Eq. (4) was obtained from Eq. (3) by the exact transformation).
Moreover, as can be verified easily, Eq. (4) presents the unitary
propagation ${\bf s}_i(t+\tau)={\bf D}_i(t,\tau) {\bf s}_i(t)$, where
${\bf D}_i(t,\tau)$ is an orthonormal matrix which rotates the vector
${\bf s}_i(t)$ on angle $\varphi=\arcsin(\tilde{\Omega}_i \tau/(1+\tau^2
\tilde{\Omega}_i^2/4))$ around axis $\tilde{\bf \Omega}_i$. Therefore,
the modified scheme will maintain the unit norm of spin lengths perfectly,
i.e. $|{\bf s}_i(t+\tau)|=|{\bf s}_i(t)|=1$, for each iteration. This
may lead to more efficient computations, despite a somewhat complicated
structure of the RHS of Eq. (4) with respect to that of Eq. (3).

The convergence of Eq. (4) can be improved significantly by using an
advanced iterative method. Namely, recalculating a current ($i=1,2,
\ldots,N$) value of ${\bf s}_i^{(n+1)}(t+\tau)$ within the $n$th
iteration, it is necessary to take into account the already obtained
quantities ${\bf s}_k^{(n+1)}(t+\tau)$ for $k=1,2,\ldots,i-1$ when
forming the RHS of Eq. (4). The advanced method is preferable for lattice
systems with the nearest-neighbor convention, when the computer time per
iteration required for the recalculations of spin values (according to
Eq. (4)) dominates over the time needed to update the mid-step Larmor
frequencies. Further optimizations are also possible for each specific
model. This completes our mid-point spin dynamics (MPSD) algorithm of
the second-order.

A way to construct higher-order versions of the MPSD integrator lies in
employing a multiple staging technique used earlier \cite{Krech} in the
framework of the ST approach. We have realized that this technique is
applicable for our approach too and obtained the following result
\begin{equation}
{\bf s}_i(t+\tau)= \prod_{p=1}^P {\bf D}_i(t,\xi_p \tau) {\bf s}_i(t)
+ {\cal O}(\tau^{m+1}) ,
\end{equation}
where the coefficients $\xi_p$ are chosen at a given number $P$ in
such a way to provide the highest possible value for $m$. The desired
fourth-order ($m=4$) algorithm (MPSD4) can be directly derived from Eq.
(5) using $P=5$ and the coefficients $\xi_1 = \xi_2 = \xi_4 = \xi_5
\equiv \xi = 1/(4 - 4^{1/3})$, and $\xi_3 = 1 - 4\xi$. That is very
interesting, these coefficients coincide with those obtained within
the ST decomposition of exponential operators \cite{Suzuki}.

Clearly, the fourth-order version is energy- and magne\-tization-%
preserving and conserves spin lengths (since the conservations are
achieved at each stage $p$). The solutions generated by the MPSD/MPSD4
algorithms are also time reversible (because past and future values
of ${\bf s}_i$ and ${\bf \Omega}_i$ enter symmetrically into the
interpolated function ${\bf \dot s}_i$, and because $\xi_p$ appear
symmetrically in Eq. (5)). The reproduction of the last feature is
particularly important as well since the numerical stability of
an algorithm is directly connected with its time reversibility
\cite{Frenkel}.

In our MD simulations we considered a simple cubic lattice in $d=3$ with
$N=1000$ sites imposing periodic boundary conditions to each direction.
The strongly anisotropic case $C=J$ with $J>0$, $J_{ij}=J \delta_{ij}$
and $\lambda=1$ was chosen to describe spin interactions (Eq. (1)),
where $\delta_{ij}=1$ for the nearest neighbor $j$ of site $i$, and
$\delta_{ij}=0$ otherwise. All test runs were started from an identical
well equilibrated configuration prepared by us with the help of MC
simulations at a temperature $T = 0.8 T_{\rm c}$, where $T_{\rm c}=
1.442929$ $J/k_{\rm B}$ is the critical temperature \cite{Chen} of the
isotropic model $(C = 0)$. The simulations were performed on the Origin
2000 workstation at the Linz University. The equations of motion were
integrated using the Adams-Bashforth-Moulton (ABM) predictor-corrector
integrator \cite{Burden} at $\tau^\ast=\tau J/\hbar=0.01$, the ST
decomposition schemes \cite{Krech} of the second (STD) and forth (STD4)
orders at $\tau^\ast=0.04$ and 0.2, respectively, as well as using our
MPSD and MPSD4 algorithms (Eqs. (4) and (5)) at $\tau^\ast=0.04$, 0.1,
0.2, and 0.4.

Examples on the total energy $E^\ast=E/J$ and magnetization $M_z$
conservations are shown in Fig.~1. The huge energy drift (see dashed
curve in Fig.~1a) indicates clearly that the ABM algorithm is
unsuitable for long-duration observations even at the smallest time
step. This is explained by the irreversibility of the ABM integrator
and the fact that it destroys the unit norm of spin lengths. At the
same time, the STD/STD4 algorithms allow much larger step sizes, that
is in the self-consistency with a conclusion of Ref. \cite{Krech}. Three
iterations were sufficient for the STD algorithm to obtain a level of
energy conservation presented in Fig.~1a. Correspondingly 5 and 6
iterations were required for the STD and STD4 algorithms to conserve
the total energy within machine accuracy ($\varepsilon \sim 10^{-9}$
in our program code). The STD/STD4 integrators, however, do not conserve
the magnetization (see Fig.~1b) which fluctuates quite visibly
especially in the STD case. These fluctuations are caused by the
${\cal O}(\tau^3)$ (or ${\cal O}(\tau^5)$) truncation errors and
will increase drastically with further increasing $\tau$.

The pattern is different for the MPSD/MPSD4 algorithms because they
can conserve both the total energy and magnetization at, in principle,
arbitrary time steps. We have determined the following values for the
number $l$ of iterations needed for obtaining the conservation to within
machine accuracy: $l=5, 8, 11$, and 18 for the MPSD as well as $l=4, 6,
7$, and 11 for the MPSD4 integrators, corresponding to the time steps
$\tau^\ast=0.04, 0.1, 0.2$, and 0.4, respectively (note that the advanced
method was used to iterate Eq. (4)). The MPSD (MPSD4) algorithm required
at $\tau^\ast=0.04$ ($\tau^\ast=0.2$) approximately the same computer
time as that of the STD (STD4) scheme for the integration over the fixed
$t^\ast= t J/\hbar=1000$ interval. However, the MPSD/MPSD4 algorithms can
be used with larger time steps ($\tau^\ast > 0.04$/$\tau^\ast > 0.2$) in
view of their energy- and magnetization-preserving properties. Taking into
account that the number $l$ increases with increasing $\tau$ slower than
linearly, these algorithms will lead to an improvement efficiency of the
computations. Of course, we cannot apply too large step sizes ($\tau^\ast
\sim 1$), because then the microscopic solutions will deviate considerably
from exact trajectories. The final decision on using the biggest possible
values of step sizes for the MPSD/MPSD4 algorithms can be done in each
specific case by direct MD measurements of macroscopic observable
quantities.

Our measurements were performed for the dynamic structure factor
$$
S(k,\omega)=\frac{1}{2\pi} \int_{-\infty}^{\infty}
\langle \sum_{i,j} {\bf s}_i^{\rm T}(0)
{\mbox{\boldmath $\cdot$}} \, {\bf s}_j^{\rm T}(t) e^{{\rm i}{\bf k}
{\mbox{\boldmath $\cdot$}}({\bf r}_i-{\bf r}_j)} \rangle e^{-{\rm i}
\omega t} {\rm d} t,
$$
which allows one to extract the spectrum of collective transverse spin
excitations, where ${\bf r}_i$ denote the lattice vectors and ${\bf s}_i^{\rm T}$
is a perpendicular to ${\bf M}$ component of ${\bf s}_i$. The function
$S(k,\omega)$ is plotted in Fig.~2a in the undimensional presentation
$S^\ast(k,\omega)=J S(k,\omega)/\hbar$ with $\omega^\ast=\omega \hbar/J$,
$k^\ast=k/k_{\rm min}$ and $k_{\rm min} = 2 \pi/N^{1/3} = \pi/5$. It
was obtained within the MPSD integration at $\tau^\ast=0.04$, and the
averaging $\langle \ \rangle$ was taken over the time $t^\ast=1000$ for
each of 1000 runs (with independent initial MC configurations). In order
to investigate the influence of increasing step size on the spectrum,
the calculations were repeated with the STD/STD4 and MPSD/MPSD4
algorithms at various step sizes.
The frequency $\omega^\ast_{\rm max}$ and the relaxation time
$\tau^\ast_{\rm cor}$ of the transverse waves, calculated numerically
at $k^\ast=1$ from the peak position and its half-width, respectively,
are shown in Fig.~2b and 2c as the functions of the time step.
As can be seen clearly, the MPSD/MPSD4 algorithms
are less sensitive to increasing $\tau$ than the STD/STD4 integrators.
For instance, the level of deviations in $\omega^\ast_{\rm max}$ obtained
with the STD (STD4) schemes at $\tau^\ast=0.04$ ($\tau^\ast=0.2$) can
be observed with the MPSD (MPSD4) algorithms at step sizes which are
approximately in factor 1.5-2.0 larger, namely, at $\tau^\ast \sim 0.07$
($\tau^\ast \sim 0.3$). This gain in time is explained by additional
cancellations of truncation errors due to the conservation of all the
integrals of motion within our approach. For the same level of accuracy,
the fourth-order versions MPSD4/STD4 allow step sizes which are nearly
in 4-5 times larger than those of the second-order algorithms MPSD/STD.
This compensates to some extent the additional computational efforts
needed to evaluate high-order expressions. However, if very high
precision is required, the fourth-order schemes become more
efficient, because then the truncation errors decrease more
rapidly with decreasing the step size.

In the conclusion we point out that alternative algorithms for classical
spin dynamics simulations have been proposed. Their advantages over the
predictor-corrector schemes are: (i) time-reversibility, (ii) exact
conservation of spin lengths, and (iii) allowance of much larger time
steps. The advantages over the decomposition integrators consists in
(i) magnetization conservation, (ii) allowance of larger step sizes,
and (iii) applicability to systems with an arbitrary lattice structure
and to continuum spin models. Moreover, the possibility of the new
algorithms to preserve all the conservation laws should be considered
as the chief feature which distinguishes them from all existing MD
integrators. There are no other algorithms of such a kind known for
any system of interacting particles. This fact may play a role in the
methodology of MD and stimulate further investigations on constructing
conservation laws preserving algorithms for other systems.

This work was supported in part by the Fonds zur F\"orderung
der wissenschaftlichen Forschung under Project No. P12422-TPH.

\newpage

\begin{center}
{\large Figure captions}
\end{center}

FIG. 1. The total energy $E^\ast/N$ (subset (a)) and magnetization $M_z/N$
(subset (b)) per spin as functions of the time length $t^\ast= t J/\hbar$
of the simulations carried out for a lattice ferromagnet model using the
predictor-corrector (dashed curve, marked as ABM), decomposition (solid
curves, STD/STD4), and mid-point (bold solid lines, MPSD) approaches.

\vspace{12pt}

FIG. 2. (a) The reduced transverse dynamic structure factor $S(k,\omega)$
of a lattice ferromagnet model; (b) and (c): The frequency peak position
(b) of $S(k^\ast,\omega)$  and the relaxation time (c) of transverse waves
as functions of the time step $\tau^\ast$, obtained at $k^\ast=1$ within
different integration schemes. Numerical errors are less than the size of
symbols used in the subsets (b) and (c).

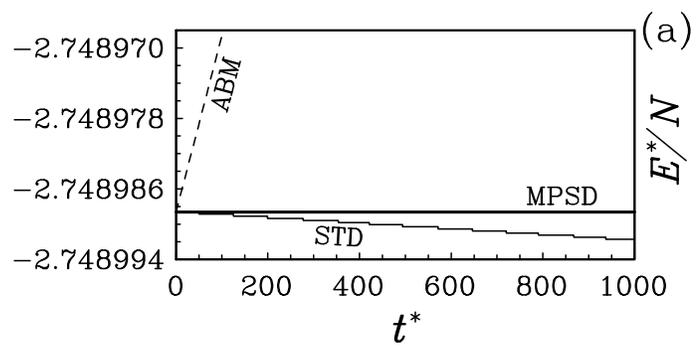
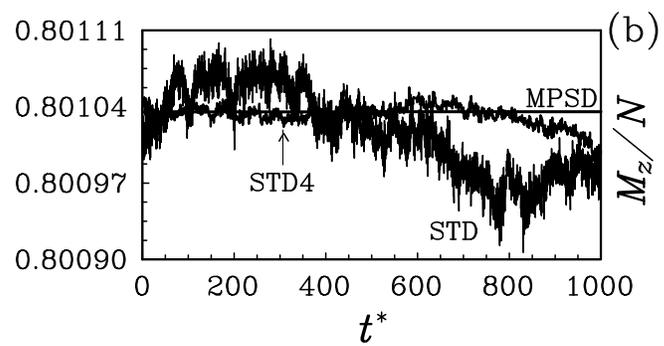
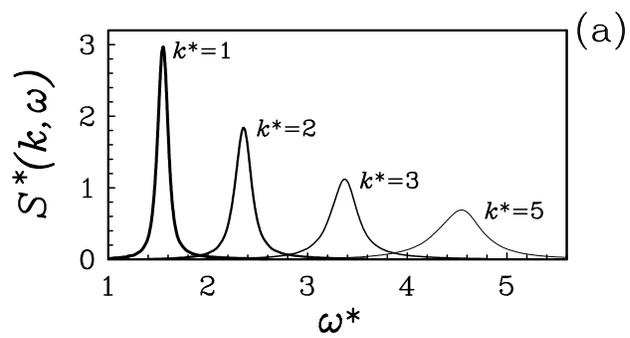
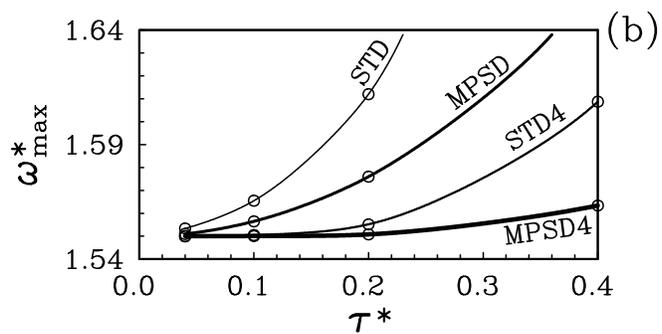
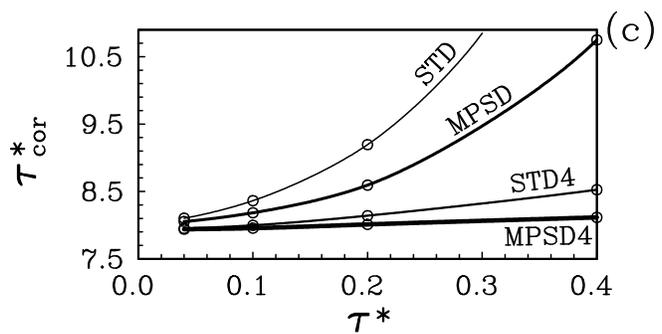
\end{document}